\documentclass[12pt]{article}
\usepackage{amsmath,amssymb}
\usepackage{bm}

\newcommand{\be}{\begin{equation}}
\newcommand{\ee}{\end{equation}}
\newcommand{\bea}{\begin{eqnarray}}
\newcommand{\eea}{\end{eqnarray}}

\title{
{\Large Reduced Trilinear Reformulation of the Nakamura Conjecture}
}
\author{
Takeshi Fukuyama\\[2mm]
{\small Research Center for Nuclear Physics (RCNP)}\\
{\small Osaka University, Ibaraki, Osaka 567--0047, Japan}
}
\date{}

\begin{document}
\maketitle
\begin{abstract}

The Tomimatsu--Sato (TS) family, characterized by the rotation parameter $q$
and the TS index $\delta=n,$ provides an important class of exact stationary axisymmetric vacuum solutions of Einstein's equations, whose integrable structure is known to be closely related to the $n$-point Toda molecule hierarchy through the Nakamura Conjecture. However, the set of equations appearing in the Nakamura Conjecture contains not only Hirota bilinear derivatives but also ordinary first-derivative terms, and therefore is not formulated entirely within the conventional bilinear algebra.

In this paper we introduce a reduced trilinear formulation based on the reduced sector $(a,b,c)\rightarrow(a,b,1)$ of the $Z_3$-symmetric trilinear Hirota operators. We show that both the Hirota bilinear derivatives and the ordinary derivatives appearing in the Nakamura Conjecture can be rewritten completely within this reduced trilinear framework. Consequently, the set of equations admits a formulation in terms of reduced trilinear operators.

We further show that the reduced trilinear formulation naturally inherits a Hirota-type direct method. The conventional bilinear spectral factor $k_i-k_j$ is replaced by the $Z_3$-weighted combinations $k_i+\omega k_j$ and $k_i+\omega^2k_j$, providing a direct-method structure characteristic of the reduced trilinear hierarchy.

These results suggest that the Toda-molecule description of the Tomimatsu--Sato hierarchy may be viewed as a reduced sector of a broader trilinear framework, and provide a new perspective on the integrable structure of stationary axisymmetric gravity.

\end{abstract}

\section{Introduction}

The stationary axisymmetric vacuum Einstein equations occupy a central
position in general relativity.  Through the Ernst formulation \cite{Ernst1968}, the field
equations are reduced to a single nonlinear complex equation, from which
many exact solutions, including the Kerr \cite{Weyl1917, Kerr1963} and Tomimatsu--Sato (TS)
families \cite{Tomimatsu1972, Tomimatsu1973, Batic2023}, have been constructed.  Despite this remarkable success, the
integrable structure underlying the higher Tomimatsu--Sato solutions has
remained much less transparent than that of many classical soliton
equations.
The TS family, characterized by the
rotation parameter $q$ and the TS index $\delta=n$, provides an important class of exact stationary
axisymmetric vacuum solutions of Einstein's equations,
whose integrable structure is known to be closely related
to the n-point Toda molecule hierarchy \cite{Toda1967, Toda1, Toda2, Toda3} through the Nakamura Conjecture \cite{Nakamura1984, Nakamura1985, Nakamura1993}.
 This conjecture was partially proved by Fukuyama, Kamimura and Yu \cite{FKY2007}, who expressed the stationary
Einstein equations in terms of Hirota bilinear equations for Toda
molecule $\tau$-functions.  For the non-rotating case ($q=0$), the
conjecture was proved for arbitrary TS index $\delta$,
while for the rotating case explicit verification has been carried out
for the lower values of $\delta$.

The Nakamura Conjecture reveals a remarkable connection between stationary
general relativity and integrable systems.  However, it also exhibits an
interesting feature.  The fundamental operator appearing in the Nakamura
Conjecture contains not only Hirota bilinear derivatives
[15] but also ordinary first derivatives, as shown in
Eqs. (1)--(5). Consequently, the hierarchy is not written
entirely within the conventional bilinear algebra.

Independently, multilinear extensions of Hirota's formalism have been
developed for nonlinear integrable systems.  In particular, the
Yu--Toda--Sasa--Fukuyama hierarchy introduced a
$Z_3$-symmetric trilinear calculus \cite{YTSF1998}, whose algebraic structure differs
essentially from the ordinary Hirota bilinear theory \cite{Grammaticos}.  The natural
question is therefore whether the Nakamura Conjecture possesses a hidden
trilinear structure.

In the present work we address this question by introducing a reduced
trilinear formulation.  Instead of the full three-slot trilinear
operators, we consider the reduced sector
$(a,b,c)\rightarrow(a,b,1)$,
which is appropriate for the stationary axisymmetric problem.  We show
that both the Hirota bilinear derivatives and the ordinary derivatives
appearing in the Nakamura Conjecture can be rewritten completely in terms of
the reduced trilinear operators.  Consequently, the Conjecture admits a formulation within a reduced trilinear algebra.

The significance of this reformulation goes beyond a mere change of
notation.  In Hirota's direct method, the bilinear representation allows
one to construct multi-soliton solutions through an exponential
$\tau$-function expansion.  We show that the same mechanism survives in
the reduced trilinear sector.  The bilinear spectral factor
$k_i-k_j$ is replaced by the $Z_3$-weighted combinations
$k_i+\omega k_j$ and $k_i+\omega^2k_j$, indicating that the reduced
trilinear hierarchy possesses its own direct-method structure.

The purpose of this paper is therefore not to replace the Toda molecule
description of the Tomimatsu--Sato hierarchy, but to demonstrate that
the Nakamura Conjecture naturally admits a reduced trilinear realization.
This provides a new viewpoint on the integrable structure of stationary
axisymmetric gravity and suggests that the Conjecture may
be regarded as a reduced sector of a broader trilinear framework.

\section{Reduced Trilinear Reformulation of the Nakamura Conjecture}

The bilinear formulation of the Nakamura Conjecture \cite{Nakamura1993} for the TS index $n$ is given by 
\begin{equation}
D_x(g_n f_n-g_n^{*}f_n^{*})=0 ,
\label{FKY6}
\end{equation}
\begin{equation}
D_y(g_n f_n+g_n^{*}f_n^{*})=0 ,
\label{FKY7}
\end{equation}
together with
\begin{equation}
F(g_n^{*}\!\cdot f_n)=0 ,
\label{FKY8}
\end{equation}
\begin{equation}
F(g_n^{*}\!\cdot g_n+f_n^{*}\!\cdot f_n)=0 ,
\label{FKY9}
\end{equation}
where
\begin{equation}
F=(x^{2}-1)D_x^{2}
+2x\partial_x
+(y^{2}-1)D_y^{2}
+2y\partial_y
+c_n .
\label{Fop}
\end{equation}

The operator $F$ contains both Hirota bilinear derivatives and
ordinary derivatives.
This suggests that the Nakamura Conjecture is not purely bilinear.

Let
\begin{equation}
\omega=e^{2\pi i/3},
\qquad
1+\omega+\omega^{2}=0 .
\end{equation}

We introduce the trilinear Hirota operators
\begin{equation}
T_x
=
\partial_{x_1}
+\omega \partial_{x_2}
+\omega^{2}\partial_{x_3},
\end{equation}
\begin{equation}
T_x^{*}
=
\partial_{x_1}
+\omega^{2}\partial_{x_2}
+\omega \partial_{x_3},
\end{equation}
and similarly for $T_y$ and $T_y^{*}$.

Restricting ourselves to the reduced sector
\begin{equation}
(a,b,c)\rightarrow(a,b,1),
\end{equation}
we obtain
\begin{equation}
T_x(a,b,1)
=
a_x b+\omega a b_x ,
\label{Txdeg}
\end{equation}
\begin{equation}
T_x^{*}(a,b,1)
=
a_x b+\omega^{2} a b_x .
\label{Txstardeg}
\end{equation}

Solving (\ref{Txdeg}) and (\ref{Txstardeg}) for the Hirota
bilinear derivative gives
\begin{equation}
D_x(a\!\cdot\! b)
=
\frac{\omega^{2}T_x(a,b,1)
-\omega T_x^{*}(a,b,1)}
{\omega^{2}-\omega}.
\label{DxTri}
\end{equation}

Similarly,
\begin{equation}
D_y(a\!\cdot\! b)
=
\frac{\omega^{2}T_y(a,b,1)
-\omega T_y^{*}(a,b,1)}
{\omega^{2}-\omega}.
\label{DyTri}
\end{equation}

The ordinary derivatives may also be rewritten:
\begin{equation}
\partial_x(ab)
=
-\omega T_x(a,b,1)
-\omega^{2}T_x^{*}(a,b,1),
\label{partialxtri}
\end{equation}
\begin{equation}
\partial_y(ab)
=
-\omega T_y(a,b,1)
-\omega^{2}T_y^{*}(a,b,1).
\label{partialytri}
\end{equation}

Equations (\ref{DxTri})--(\ref{partialytri})
show that both the Hirota derivatives and the ordinary derivatives
appearing in the Nakamura Conjecture can be expressed within the
reduced trilinear sector $(a,b,1)$.

To rewrite these equations in the reduced trilinear language,
we first note that
\begin{equation}
D_x^2(a\!\cdot\! b)
=
a_{xx}b
-2a_xb_x
+ab_{xx},
\end{equation}
and similarly for $D_y^2$.

Using the reduced trilinear operators,
the second-order Hirota derivative may be expressed as
\begin{equation}
D_x^2(a\!\cdot\! b)
=
-\frac{\omega^2}{3}
T_x^2(a,b,1)
-\frac{\omega}{3}
T_x^{*2}(a,b,1)
+\frac{2}{3}
T_xT_x^{*}(a,b,1),
\label{Dx2Tri}
\end{equation}
and
\begin{equation}
D_y^2(a\!\cdot\! b)
=
-\frac{\omega^2}{3}
T_y^2(a,b,1)
-\frac{\omega}{3}
T_y^{*2}(a,b,1)
+\frac{2}{3}
T_yT_y^{*}(a,b,1).
\label{Dy2Tri}
\end{equation}

Substituting (\ref{Dx2Tri}),
(\ref{Dy2Tri}),
(\ref{partialxtri}),
and (\ref{partialytri})
into (\ref{Fop}),
the operator $F$ can be written entirely
in terms of reduced trilinear operators:

\begin{align}
{\cal F}_T(a,b,1)
={}&
(x^2-1)
\left[
-\frac{\omega^2}{3}T_x^2
-\frac{\omega}{3}T_x^{*2}
+\frac{2}{3}T_xT_x^{*}
\right](a,b,1)
\nonumber\\
&
+(y^2-1)
\left[
-\frac{\omega^2}{3}T_y^2
-\frac{\omega}{3}T_y^{*2}
+\frac{2}{3}T_yT_y^{*}
\right](a,b,1)
\nonumber\\
&
-2x
\left[
\omega T_x
+\omega^2T_x^{*}
\right](a,b,1)
\nonumber\\
&
-2y
\left[
\omega T_y
+\omega^2T_y^{*}
\right](a,b,1)
+c_n\,ab .
\label{FToperator}
\end{align}

Consequently, equation (\ref{FKY8}) becomes

\begin{equation}
{\cal F}_T(g_n^{*},f_n,1)=0 .
\label{TriFKY8}
\end{equation}

Similarly, equation (\ref{FKY9}) becomes

\begin{equation}
{\cal F}_T(g_n^{*},g_n,1)
+
{\cal F}_T(f_n^{*},f_n,1)
=0 .
\label{TriFKY9}
\end{equation}

Equations (\ref{TriFKY8}) and (\ref{TriFKY9})
show that the complete Nakamura Conjecture admits a formulation
within a reduced trilinear algebra.

\section{Reduced Trilinear Expansion and Soliton Ansatz}
In Hirota's direct method, the significance of a
bilinear equation is not merely that it is bilinear.
Its essential feature is that the exponential
perturbation expansion terminates after a finite number
of orders, thereby yielding exact multi-soliton
solutions (see Sec.1.7 of Ref.~\cite{Hirota2004} for details).
One may therefore ask whether an analogous
algebraic mechanism survives in the reduced
trilinear sector. Motivated by Hirota's
construction, we consider the exponential
expansion
\begin{equation}
\tau
=
1+\epsilon e^{\eta_1}
+\epsilon^2 A_{12}e^{\eta_1+\eta_2}
+\cdots ,
\label{HirotaExpansion}
\end{equation}
where
\begin{equation}
\eta_i=k_i x+\ell_i y+\eta_i^{(0)} ,
\end{equation}
and determine the dispersion relations and interaction coefficients
order by order in $\epsilon$.

For a bilinear Hirota derivative one has
\begin{equation}
D_x(e^{\eta_i}\cdot e^{\eta_j})
=
(k_i-k_j)e^{\eta_i+\eta_j}.
\label{BilinearDispersion}
\end{equation}
Thus the bilinear equation is converted into an algebraic equation for
the wave numbers.
The same algebraic mechanism survives
in the reduced trilinear sector.
Indeed, for the trilinear Hirota operator
\begin{equation}
T_x
=
\partial_{x_1}
+\omega\partial_{x_2}
+\omega^2\partial_{x_3},
\qquad
1+\omega+\omega^2=0,
\end{equation}
one obtains
\begin{equation}
T_x(e^{\eta_i},e^{\eta_j},1)
=
(k_i+\omega k_j)e^{\eta_i+\eta_j}.
\label{ReducedTrilinearDispersion}
\end{equation}
Similarly,
\begin{equation}
T_x^*(e^{\eta_i},e^{\eta_j},1)
=
(k_i+\omega^2 k_j)e^{\eta_i+\eta_j}.
\label{ReducedTrilinearDispersionStar}
\end{equation}

Therefore, after the Nakamura Conjecture is rewritten in the reduced
trilinear language, the same direct-method logic applies.
The role of the bilinear factor $k_i-k_j$ is replaced by the
$Z_3$-weighted factors
\begin{equation}
k_i+\omega k_j,
\qquad
k_i+\omega^2 k_j .
\end{equation}

Consequently, a reduced trilinear equation of the form
\begin{equation}
P(T_x,T_y,T_x^*,T_y^*)\,\tau\cdot\tau\cdot 1=0
\end{equation}
leads, under the exponential ansatz, to an algebraic dispersion
condition
\begin{equation}
P(k_i+\omega k_j,\ell_i+\omega \ell_j,
  k_i+\omega^2 k_j,\ell_i+\omega^2 \ell_j)=0 .
\end{equation}

This is the precise sense in which the reduced trilinear formulation is
not merely a rewriting of the bilinear formalism.  It provides a
Hirota-type direct method in which the elementary spectral factors are
not differences of wave numbers, but $Z_3$-weighted combinations.

In the present gravitational problem, the third slot is frozen as
$(a,b,1)$ because of the stationary axisymmetric reduction.  Nevertheless,
the reduced trilinear structure retains enough of the original
three-slot algebra to support a Hirota-type soliton expansion.
Thus, if the Toda--molecule hierarchy underlying the Tomimatsu--Sato
solutions is expressed in reduced trilinear form, one may in principle
construct its solutions by following the same order-by-order procedure
as in Hirota's bilinear method, with the replacement
\begin{equation}
k_i-k_j
\quad\longrightarrow\quad
k_i+\omega k_j,\quad k_i+\omega^2 k_j .
\end{equation}
\section{Conclusion}

In this paper we have presented a reduced trilinear reformulation of the Nakamura Conjecture underlying the
Tomimatsu--Sato solutions of the stationary axisymmetric vacuum
Einstein equations.

The essential observation is that not only the Hirota bilinear
derivatives but also the ordinary first-derivative terms appearing in
the Nakamura Conjecture can be rewritten completely in terms of reduced
trilinear Hirota operators acting on the reduced sector
$(a,b,1)$.  Consequently, the complete Nakamura Conjecture admits a
description within a reduced trilinear algebra.

We have further shown that the significance of this reformulation is not
merely algebraic.  Following Hirota's direct method, the reduced
trilinear operators naturally lead to a modified spectral structure in
which the bilinear factor $(
k_i-k_j
)$
is replaced by the $Z_3$-weighted combinations
$(
k_i+\omega k_j
)$
and
$(
k_i+\omega^2k_j
)$.
This indicates that the reduced trilinear formulation possesses a
direct-method structure analogous to the conventional Hirota bilinear
theory.

The present work should therefore be regarded as a first step toward a
multilinear understanding of the integrable structure of the
Tomimatsu--Sato hierarchy.  At the present stage, the third slot of the
trilinear operator is frozen by the stationary axisymmetric reduction,
and the resulting structure is a reduced trilinear hierarchy rather than
a genuine three-slot multilinear system.  Nevertheless, the appearance
of the characteristic $Z_3$ algebra suggests that the conventional Toda
molecule description may represent only a reduced sector of a broader
multilinear framework.

Several important questions remain open. It would be desirable to clarify whether the determinant identities underlying the Toda molecule hierarchy admit an intrinsic reduced trilinear formulation, and whether the present construction can be extended from the reduced sector (a,b,1) to a genuine three-slot trilinear hierarchy. Such an extension may provide not merely a generalization of the Toda-molecule description, but a new framework for constructing and classifying broader classes of exact stationary axisymmetric black-hole solutions beyond the present Tomimatsu–Sato family.


\end{document}